# Domain wall dynamics in stepped magnetic nanowire with perpendicular magnetic anisotropy

S. Al Risi[1], R. Sbiaa[1]* and M. Al Bahri[2]

[1]Department of Physics, Sultan Qaboos University, P.O. Box 36, PC 123, Muscat, Oman
[2] Department of Basic Sciences, A'Sharqiyah University, Post Box 42, PC 400, Ibra, Oman

Corresponding Email id: * rachid@squ.edu.om

**Abstract**

Micromagnetic simulation is carried out to investigate the current-driven domain wall (DW) in a nanowire with perpendicular magnetic anisotropy (PMA). A stepped nanowire is proposed to pin DW and achieve high information storage capacity based on multi-bit per cell scheme. The DW speed is found to increase for thicker and narrower nanowires. For depinning DW from the stepped region, the current density $J_\text{dep}$ is investigated with emphasis on device geometry and materials intrinsic properties. The $J_\text{dep}$ could be analytically determined as a function of the nanocontriction dimension and the thickness of the nanowire. Furthermore, $J_\text{dep}$ is found to exponential dependent on the anisotropy energy and saturation magnetization, offering thus more flexibility in adjusting the writing current for memory applications.

# 1. Introduction

Magnetic random access memory (MRAM) gained attention as a possible replacement to conventional memories such as static and dynamic RAMs and even flash memory. It has several advantages as reported in few articles [1–6]. For the MRAM device based on magnetic tunnel junctions (MTJ)s, the direction of magnetization defines the magnetic state, thus only one bit per cell could be stored. For increasing the storage capacity of the memory, there were attempts to use MTJs with two free layers to achieve four magnetic states [7,8]. For even more storage capacity, domain wall (DW)-based MRAM was intensively investigated [9–21]. The first generation of DW-based devices was based on materials with in-plane magnetic anisotropy (IMA). However, the continuous reduction of device size makes these materials less effective in keeping the magnetic state (information) stable. On the other hand, materials with perpendicular magnetic anisotropy (PMA) have the advantage to provide better thermal stability due to the higher magnetic anisotropy.

Different studies have been focused on domain wall movement in nanowires with in-plane anisotropy and an emphasis on the dependence of the velocity of DW on nanowire dimensions. It has been reported that DW motion can be controlled by adjusting the nanowire width and thickness [22,23]. For in-plane (IMA) material, DW velocity could be made larger by increasing nanowire width and decreasing its thickness.

To store the information in magnetic domains, the DW should be stable at a predefined position within the nanowire. Pinning DW at a precise position has been investigated with several schemes [26-27]. One of them was based on creating triangular notches with different geometries for DW made of permalloy material [24-27]. Another way like using a magnetic field perpendicular to the direction of DW motion to reduce the speed of DW to zero at a required position was also proposed [28]. In such a method, the magnetic field needed to pin DW is higher than the Walker field and causes deformation for the DW structure.

In our previous work, we proposed nanowire with a stepped area to pin DW in materials with in-plane magnetic anisotropy [23,29]. In this paper, we focus on material with perpendicular anisotropy and we used micromagnetic formalism with the inclusion of spin-transfer torque effect with both adiabatic and non-adiabatic terms.

# 2. Theoretical model

In this study, we investigate a magnetic nanowire of fixed length L = 200 nm and varied width $w$ and thickness $t_z$ as shown in Fig.1(a). The device is divided into small meshes with the size of $(2.5 \times 2.5 \times t_z)$ nm$^3$. The size of the mesh is chosen to be smaller than the exchange length $l_{ex}$ of the materials considered in this study. In the last part, the saturation magnetization ($M_s$), the perpendicular magnetic anisotropy ($K_u$) are varied while the exchange length ($A$) was kept constant ($A = 2 \times 10^{11}$ J/m). These chosen values correspond to the actual magnetic properties of the Co/Ni multilayers as reported previously [30]. The calculation is carried out with the object-oriented micromagnetic framework (OOMMF) based on the Landau-Lifshitz-Gilbert (LLG) equation with spin-transfer torque (STT) term [31].

$$\frac{d\mathbf{m}}{dt} = -\gamma(\mathbf{H}_{\text{eff}} \times \mathbf{m}) + \alpha\left(\mathbf{m} \times \frac{d\mathbf{m}}{dt}\right)$$
$$+ \mathbf{u}.\times\left(\mathbf{m} \times \frac{\partial \mathbf{m}}{\partial t}\right) + \beta.\mathbf{u}.\mathbf{m} \times \frac{\partial \mathbf{m}}{\partial t} \quad (1)$$

where $\mathbf{m}$ is the normalized magnetization vector, $\gamma$ is the gyromagnetic ratio, $\alpha$ is the Gilbert damping parameter, the vector $\mathbf{u} = (gP\mu_B/2eM_s)\mathbf{J}$ is the adiabatic spin torque which has the dimension of velocity, $J$ is the current density, $g$ is the Lande factor, $P$ is the spin polarization, $e$ is the charge of electron and $\mu_B$ is the Bohr magneton. In the middle of the nanowire, a stepped junction is designed by shifting one part in the $x$-direction and the other in the $y$-direction, called $\lambda$ and $d$, respectively to pin DW [Fig.1 (b)]. Initially, the magnetic moments are saturated in $z$-direction then the domain wall was created at the left edge of the nanowire so the DW could move easily under a reasonably low current density. The DW could be pinned at the stepped region, then at a certain value of current density, the DW is released. This critical current density is called the depinning current density ($J_{dep}$). The calculations are performed by using LLG equation. Different values of the width and thickness of the nanowire were chosen to explore the dependence of DW velocity and the depinning current on device geometry. The dimensions of the stepped area is also varied. Fig. 1(a) shows the conventional nanowire while Fig. 1(b) illustrates the proposed stepped type which is efficient in pinning/ stabilizing DW for memory applications.

## 3. Results and Discussion

In this study, all magnetic moments within the

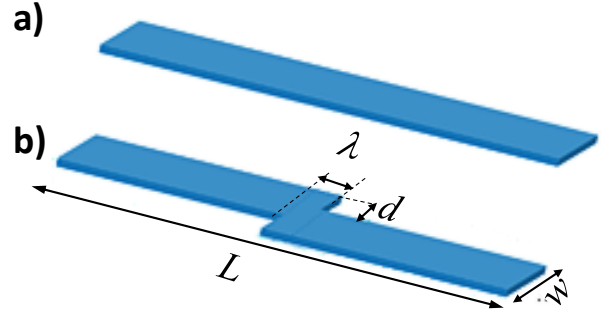

**Figure 1.** (a) Conventional nanowire with length L and width $w$ and (b) Nanowire with a stepped area at its center with dimensions $d$ and $\lambda$ to pin the DW.

nanowire are initially considered aligned in the negative $z$-direction (perpendicular to the nanowire plane). The spin-polarized current was then applied from the left edge of the nanowire to the right edge along the $x$-direction perpendicular to the direction of magnetization. Firstly, the simulations are performed using a conventional nanowire without any stepped region to investigate the DW dynamics as shown in Fig. 1 (a). In this case, DW is created as an initial state at nanowire left edge and moves through the nanowire without observing any pinning within the device as shown in Fig.2 (a). The motion of DW from the initial state ($m_z = -1$) to the final state ($m_z = +1$) happens with constant velocity as shown by the linear dependence of $m_z$ with time ($m_z = M_z/M_s$ is the normalized $z$-component of the magnetization). The insets of Fig. 1(a) are captured images showing a temporaire displacement of DW along the nanowire. Fig. 2(b) shows the plots of DW velocity $v$ as a function of $J$ for different widths of the nanowire. It is found that $v$ has a linear dependence with $J$, similar to what was reported by several groups for nanowires with

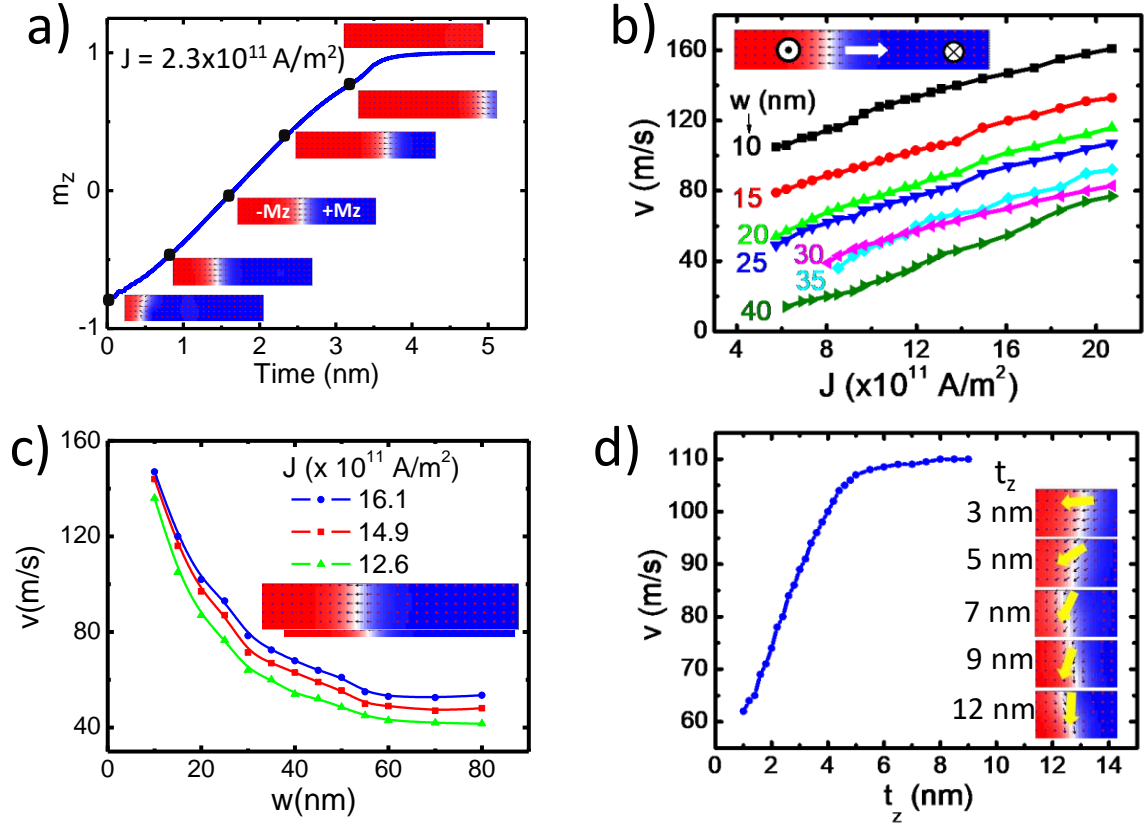

Fig. 2. (a) Normalized magnetization of a conventional nanowire as a function of time for device with L=200 nm, w = 40 nm, $M_s = 802 \times 10^3$ A/m, $K_u = 5.3 \times 10^5$ J/m$^3$, $A = 2 \times 10^{11}$ J/m and $J = 23 \times 10^{11}$ A/m$^2$ (b) plot of the DW average velocity as a function of current density for various nanowire widths. (c) DW average velocity as a function of the width of conventional nanowire for various applied current density values and (d) a plot of the average velocity and the conventional nanowire thickness for $J = 23 \times 10^{11}$ A/m$^2$. Insets are snapshots of magnetic moments at the vicinity of DW showing the transformation from Néel wall to Bloch wall.

PMA [13,32–35]. The velocity versus current density could be described by

$$v = \frac{\varepsilon \gamma \hbar P}{2eM_s\alpha} J \qquad (2)$$

where $\hbar$ is the reduced Planck constant and $\varepsilon$ is the non-adiabatic parameter [32,36].

For a device with 10 nm width, the velocity rose from 110 m/s (at a current density of $6 \times 10^{11}$ A/m$^2$) to 160 m/s (at a current density of $20 \times 10^{11}$ A/m$^2$). We confirmed this result by investigating the influence of nanowire width on the dynamics of DW. From Fig. 2(c), it could be seen that the DW velocity increases as nanowire is made narrow following the relation $J = i/w\Delta$. The equation (2) becomes,

$$v = \frac{\varepsilon \gamma \hbar P i}{2eM_s\alpha \, w\Delta} \qquad (3)$$

where $i$ the current flows, $w$ is the nanowire width and $\Delta$ is the DW width [35].

It has been argued that as the nanowire width becomes smaller, the hard axis anisotropy field decrease and the Néel wall becomes more stable, therefore the current density enhances the DW dynamics [37]. The other key parameter which has a strong effect on DW dynamics is the nanowire thickness. Fig. 2(d) is a plot of the dependence of $v$ on $t_z$. Insets are the domain wall configuration for different nanowires thicknesses it is found that the velocity has an almost linear dependence on the thickness for a certain range of values ($t_z$ between 1 and 6 nm) and agrees with the relation,

$$v = \frac{0.01 \gamma a_J t_z}{\alpha} \qquad (4)$$

where $a_J = \hbar JP/2t_z e M_s$ [36]. In this study, a current density of $23 \times 10^{11}$ A/m$^2$ was applied and $t_z$ was varied from 1 nm to 12 nm while the other parameters were kept constant. It can be seen that DW velocity increases rapidly as thickness varied from 1 nm to 5 nm to reach a velocity of about 110 m/s for a device with 5 nm thickness. It can also be seen from Fig. 2 (d) that the velocity did not change much for large thickness (more than 5 nm). As the thickness varies, the micromagnetic simulation shows the transformation from a Néel wall to Bloch wall [insets of Fig. 2 (d)]. This could be the reason for the fast increase of the velocity with thickness.

For the use of the stepped nanowire as a memory device, the effect of the stepped region dimension of the depinning current density is of great importance. There is a balance between the stability of the domain wall and the minimum $J_{dep}$ needed to move it to the next step (i.e., state). Firstly, the DW is created in the left edge of stepped nanowire and the current is applied to move the DW. Once the DW is stabilized/pinned in the stepped region, considered as the initial state, the current was varied continuously until it is depinned. The depinning current density was investigated for stepped nanowires with widths of values changing from 20 nm to 60 nm with a step of 0 nm while the length of the step ($d$) was fixed to ($w/2$). From Fig. 3 (a), it can be seen that for $w$ = 20, 30 and 40 nm, $J_{dep}$ decreases exponentially with $\lambda$. For larger values of $w$ an almost linear behavior was observed. The geometry of the stepped region provides an easy way to adjust the pinning strength for stabilizing DW at a defined position. By enlarging its lateral size ($\lambda$), DW can be depinned more easily. Fig. 3(b) shows the dependence of $J_{dep}$ on the thickness of the device for different values of $d$. The most striking result to emerge from the data is that $J_c$ increases with the thickness for different values of $d$ (here $\lambda$ was fixed to zero). Hence with these interesting results, we can achieve a higher depinning current and higher speed with larger thickness and narrow nanowire. The numerical data obtained from LLG equation can be described analytical using the expression $J_{dep} = A + Be^{-\lambda/\xi}$. Here $\lambda$ is the length of nano-constriction in the $x$-direction and $\xi$ is a fitting parameter as well as $A$ and $B$. The values of $\xi$ for different nanowire widths $w$ are summarized in Tab. 1. It is important to mention that for small values of $w$, $J_{dep}$ shows a linear behavior with $\lambda$ as can be seen in Fig. 3(a). By looking carefully at Fig. 3(b), it can be seen that $J_{dep}$ has an exponential growth with the nanowire thickness. In this case, the depinning current density can be expressed as

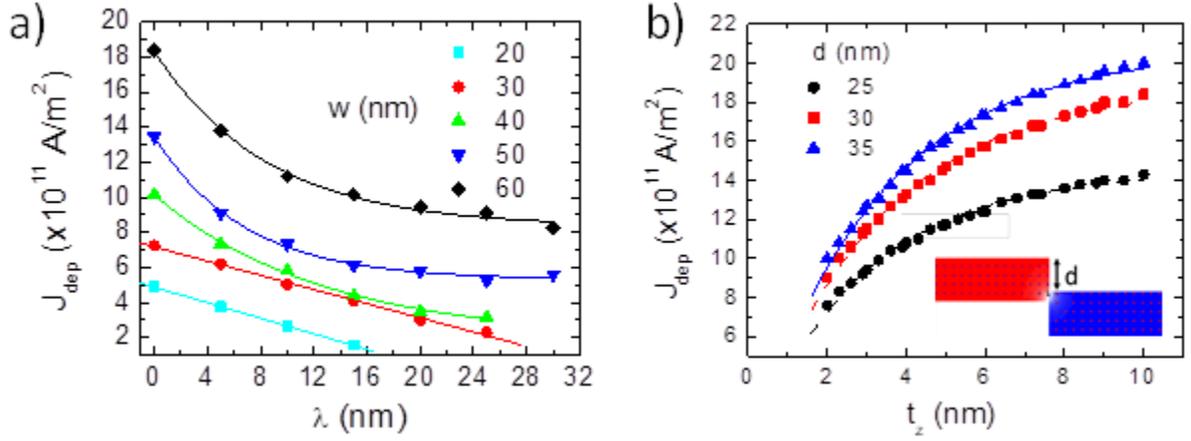

Figure 3. (a) Plots of $J_{dep}$ versus the step width $\lambda$ for various widths of nanowire $w$. For the plots, $d$ was fixed to $w/2$. (b) $J_{dep}$ versus the nanowire thickness $t_z$ for $d = 25$, 30 and 35 nm and $\lambda = 0$.

$J = J_0(1 - e^{-\rho t_z})$. In the same manner as in Fig. 3(a), the parameters $J_0$ and $\rho$, which are reported in Tab. 2, could be obtained for different values of $d$ (the nanoconstriction size along $y$-direction). From table 2, one could see that the fitting parameter $\rho$ is not much dependent on $d$ in the investigated range. The parameter $J_0$, which is considered as the current density for relatively thick nanowire, is continuously increasing with $d$ as expected. In fact, $J_{dep}$ is higher for large values of $d$ for any nanowire thickness value [Fig. 3(b)]. After investigating the effect of device geometry on the dynamics of DW and the depinning current density, the study focused on the effect of the key material properties such as $M_s$ and $K_u$ on $J_{dep}$. Firstly, $M_s$ was fixed while $K_u$ was varied between $5.0 \times 10^5$ and $10.0 \times 10^5$ J/m$^3$. In this calculation, DW was first brought to the stepped region then current was applied to depin it from its initial position. Fig. 4(a) is a plot of $J_{dep}$ as a function of $K_u$ for different values of $d$ ($\lambda$ was fixed to 0 nm). Here, the saturation magnetization was fixed to 800kA/m. It was found that the values of $J_{dep}$ obtained from numerical calculation using LLG equation could be fitted to an exponential growth function $\exp(K_u/t_1)$. The parameter $t_1$ is a fitting parameter that has the dimension of energy. From the best fit, $t_1$ is found to be almost constant for $d = 30$ and 35 nm with values of $6.9\times10^5$ and $7.0\times10^5$ J/m$^3$, respectively. A slightly larger drop of $t_1$ to $4.5\times10^5$ J/m$^3$ was obtained for $d = 25$ nm. Similarly, the effect of $M_s$ on $J_{dep}$ was investigated. It can be seen from Fig. 4(b) that under the same device geometry, $J_{dep}$ shows, in contrast, and exponential decay (1$^{st}$ order) with $M_s$ for a constant $K_u$. The plots in Fig. 4(b) are for $K_u = 5.3 \times 10^5$ J/m$^3$. The $J_{dep}$ can be described by the function $\exp(-M_s/t_2)$ where $t_2$ is a fitting parameter with values of 345, 656 and 1166 kA/m for $d = 25$, 30 and 35 nm, respectively. Optimizing $J_{dep}$ is essential for memory applications. It should be reasonably low in order to move the DW from one step to the other with a low applied current, i.e., changing the magnetic state by displacing DW within the nanowire. One can design the nanoconstriction dimensions for different values of $K_u$ and $M_s$ as described analytically above by the growth and the decay functions, respectively.

Tab. 1. The dependence of the fitting parameters $\xi$ and $B$ on the width of the nanowire. The device dimensions and materials properties as taken from Fig. 3(a)

| $w$ (nm) | $\xi$ (nm) | $B(\times 10^{11}$ A/m$^2)$ |
|---|---|---|
| 40 | 12.6 ± 1.0 | 8.2 ± 0.3 |
| 50 | 6.8 ± 0.5 | 8.1 ± 0.2 |
| 60 | 8.5 ± 0.7 | 10.0 ± 0.3 |

Tab. 2. The fitting parameters taken from exponential growth function and plotted in Fig. 3(b) for different values of $d$.

| $d$ (nm) | $\rho$ (nm$^{-1}$) | $J_0(\times 10^{11}$ A/m$^2)$ |
|---|---|---|
| 25 | 0.350 ± 0.006 | 14.5 ± 0.1 |
| 30 | 0.307 ± 0.005 | 18.9 ± 0.1 |
| 35 | 0.311 ± 0.004 | 20.6 ± 0.1 |

For the evaluation of the influence of $K_u$ on DW stability and its magnetic configuration in the vicinity of the stepped area, we conducted the calculation of time dependence of the magnetization for three values of $K_u$. Figures 5(a) and (b) show snapshot images of the nanowire with $K_u = 5\times10^5$ J/m$^3$ and $10 \times 10^5$ J/m$^3$ ($M_s$ was fixed to 800 kA/m). These are typical values for Co/Pt, Co/Ni, Co/Ni/Co/Pt multilayers and CoFeB commonly used for magnetic materials with perpendicular anisotropy [30,38–43]. The device dimensions including the nanoconstriction size are L = 200 nm, w = 40 nm, $d = 35$ nm and $\lambda = 10$ nm. The calculation was carried with a current density of $4.1\times10^{11}$ A/m$^2$. Although DW could be stabilized in the stepped area, it will not remain for a long time. Fig. 5(c), shows that DW takes less than 5 ns before vanishing; i.e., moving to the right edge of the device. To overcome this issue, an increase of $K_u$ is necessary, which has been demonstrated for $K_u = 10\times10^5$ J/m$^3$.

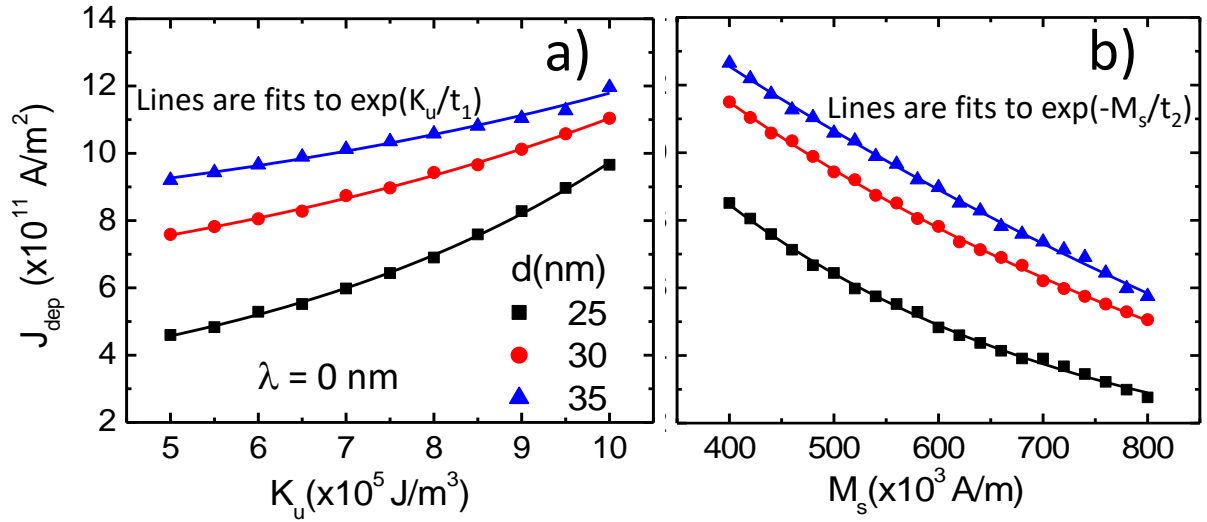

**Figure 4.** Domain wall depinning current density versus (a) $K_u$ and (b) $M_s$ for different values of $d$. The lines are fit to exponential growth function for case (a) and exponential decay for case (b). The nanowire length and width are fixed to 200 nm and 40 nm, respectively.

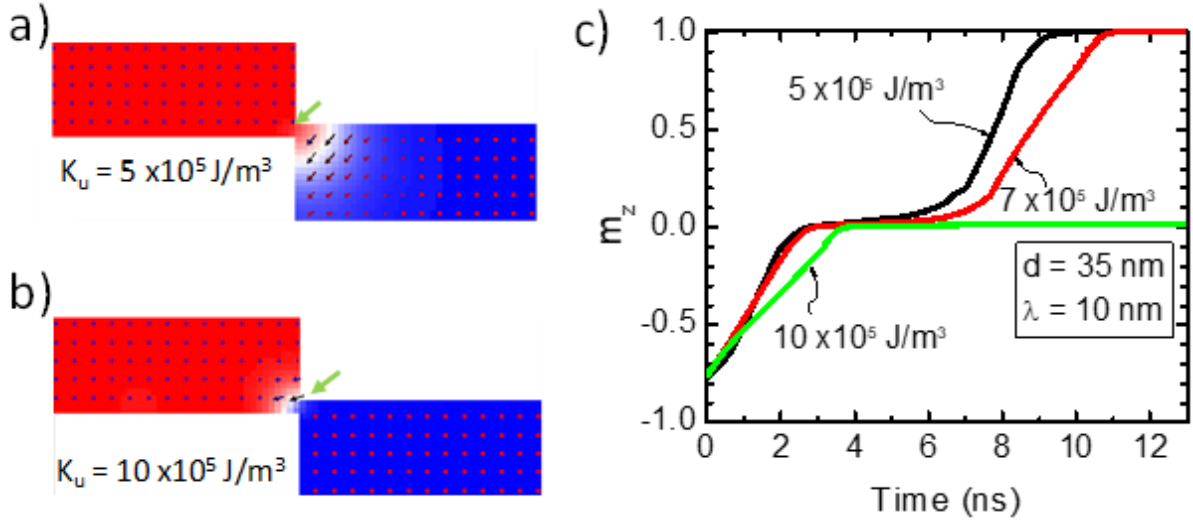

**Figure 5.** (a) Snapshot images of the nanowire near the stepped area for $K_u = 5.0 \times 10^5$ J/m$^3$ and $5.5 \times 10^5$ J/m$^3$. The length and width of the nanowire are fixed to 200 nm and 40 nm, respectively and the stepped region dimensions $d$ and $\lambda$ are fixed to 35 nm and 10 nm, respectively. The calculation is conducted for $M_s = 800$ kA/m and $J = 4.1 \times 10^{11}$ A/m$^2$. (b) Normalized magnetization in $z$-direction versus time for three values of $K_u$ ($4.5 \times 10^5$, $5.0 \times 10^5$ and $5.5 \times 10^5$ J/m$^3$).

It is important to mention that it is also possible to stabilize DW at the stepped region for relatively low $K_u$ by enlarging $d$ or shrinking $\lambda$. We noticed that magnetic moments configuration near the stepped area changes with $K_u$ (Fig. 5).

The dynamics of DW at the stepped area based on the influence of $M_s$ was also investigated. In the same manner, DW could be stabilized at the nanoconstriction for small values of $M_s$.

The stepped type nanowire offers an easy and accurate way of pinning DW within the device. In addition to the nanoconstriction dimension, the material magnetic proerties, such as $M_s$ and $K_u$, provide a further degree of freeomds to lower the depinning current density and stabilizing the DW for high capacity memory applications.

## 4. Conclusion

Domain wall dynamics was investigated in magnetic nanowires with perpendicular anisotropy. A comparison between conventional and stepped nanowires was carried out. In the first scheme, the velocity showed a linear increase with the current density and dropped exponentially with the device width. In the stepped nanowire scheme, it was found that for particular dimensions of the nanoconstriction region, DW could be stabilized and pinned. The dependence of depinning current density $J_{dep}$ on the thickness of the nanowire and dimensions of the stepped region could be described analytically. Furthermore, the dependence of $J_{dep}$ on the material properties was studied. The $J_{dep}$ could also be fitted analytically by either an exponential growth or exponential decay functions with $K_u$ and $M_s$, respectively. Tuning and predicting analytically the device geometry and materials properties for an optimal $J_{dep}$ is important for designing DW-based memory devices.

## References

[1] Wolf S. A., Awschalom D. D., Buhrman R. A., Daughton J. M., Molnár S. von, Roukes M. L., Chtchelkanova A. Y., Treger D. M. 2001 Science **294** 1488.

[2] Žutić J., Fabian J., Sarma S.D. 2004 Reviews of modern physics **76** 323.

[3] Ikeda S., Hayakawa J., Lee Y. M., Matsukura F., Ohno Y., Hanyu T. 2007 IEEE Trans. Electron Devices **54** 991.

[4] Fert A. 2008 Rev. Mod. Phys **80** 41517.

[5] Kent A., Worledge A. 2015 Nat. Nanotechnol **10** 187.

[6] Sbiaa R., Piramanayagam S. N. 2017 Phys. Status Solidi-RRL **11** 1700163.

[7] Sbiaa R., Law R., Lua S. Y. H., Tan E. L., Tahmasebi T., Wang C. C., Piramanayagam S. N. 2011 Appl. Phys. Lett. **99** 092506.

[8] Sbiaa R., 2015 J. Phys. D: Appl. Phys **48** 195001.

[9] Beach G. S. D., Nistor C., Knutson C., Tsoi M., Erskine J.L. 2005 Nature Mater **4** 741.

[10] Jubert P.-O., Klaui M., Bischof A., Rudiger U. Allenspach R. 2006 J. Appl. Phys. **99** 08G523.

[11] Parkin S. S. P., Hayashi M. Thomas L. 2008 Science **320** 190.

[12] O'Brien L., Read D. E., Zeng H. T., Lewis E. R., Petit D., Cowburn R. P. 2009 Appl. Phys. Lett. **95** 232502.

[13] Moore T. A., Miron I. M., Gaudin G., Serret G., Auffret S., Rodmacq B., Schuhl A., Pizzini S., Vogel J., Bonfim M. 2008 Appl. Phys. Lett. **93** 262504.

[14] Kim S.-R, Lee J.-Y., Choi Y.-S., Yu K. Guslienko, Lee K-S. 2008 Appl. Phys. Lett. **93** 052503.

[15] Heyne L., Rhensius J., Bisig A., Krzyk S., Punke P., Kläui M., Heyderman L. J., Guyader L., Nolting F. 2010 Appl. Phys. Lett. **96** 032504.

[16] Lahtinen T.H.E., Franke K.J.A., Van Dijken S. 2011 Sci. Rep. **2** 258.

[17] Ueda K., Koyama T., Hiramatsu R., Chiba D., Fukami S., Tanigawa H., Suzuki T., Ohshima N., Ishiwata N., Nakatani Y., Kobayashi K., Ono T. 2012 Appl. Phys. Lett. **100** 202407.

[19] Sbiaa R., Piramanayagam S. N. 2014 Appl. Phys. A **114** 1347.

[20] Hayward T. J. 2015 Sci. Rep. **5** 13279.

[21] Emori S., Umachi C. K., Bono D. C., Beach G. S. D. 2015 J. Magn. Magn. Mat. **378** 98.

[22] Ai J., Miao B., Sun L., You B., Hu A., Ding H. 2011 J. Appl. Phys. **110** 093913.

[23] Al Bahri M., Sbiaa R. 2016 Sci. Rep. **6** 28590.

[24] Hayashi M., Thomas L., Rettner C., Moriya R., Jiang X., Parkin S. 2006 Phys. Rev. Lett. **97** 1.

[25] Martinez E., Lopez-Diaz L., Torres L., Tristan C., Alejos O. 2007 Phys. Rev. B - Condens. Mater. Phys. **75** 1.

[26] Noh S. J., Miyamoto Y., Okuda M., Hayashi N., Keun Kim Y. 2012 J. Appl. Phys. **111** 1.

[27] Kunz A., Priem J. D. 2010 IEEE Trans. Magn. **46** 1559.

[28] Ermolaeva O. L., Skorokhodov E. V., Mironov V. L. 2016 Phys. Solid State **58** 2223.

[29] Al Bahri M., Borie B., Jin T. L., Sbiaa R., Klaui M., Piramanayagam S. N. 2019 Phys. Rev. Appl. **11** 024023.

[30] Al Subhi A., Sbiaa R. 2019 J. Magn. Magn. Mater. **489** 165460.

[31] Thiaville A., Nakatani Y., Miltat J., Suzuki Y., 2005 EPL **69** 990.

[32] Noh S. J., Tan R. P., Chun B. S. Kim Y. K. J. 2010 Magn. Magn. Mater. **322** 3601.

[33] Fukami S., Suzuki T., Nakatani Y., Ishiwata N., Yamanouchi M., Ikeda S., Kasai K., Ohno H. 2011 Appl. Phys. Lett. **98** 082504.

[34] Koyama T., Chiba D., Ueda K., Tanigawa H., Fukami S., Suzuki T., Ohshima N., Ishiwata N., Nakatani Y., Ono T. 2011 Appl. Phys. Lett. **98** 192509.
[35] Sethi P., Krishnia S., Gan W., Kholid F., Tan F., Maddu R., Lew W. 2017 Sci. Rep. **7** 4964.
[36] Ngo D. T., Ikeda K., Awano H. 2011 Appl. Phys. Express **4** 093002.
[37] Khvalkovskiy A. V., Zvezdin K. A., Gorbunov Y. V., Cros V., Grollier J., Fert A., Zvezdin A. K. 2009 Phys. Rev. Lett. **102** 067206.
[38] Girod S., Gottwald M., Andrieu S., Mangin S., McCord J., Fullerton E. E., Beaujour J.-M. L., Krishnatreya B. J., Kent A. D. 2009 Appl. Phys. Lett. **94** 262504.
[39] Bandiera S., Sousa R. C., Rodmacq B., Dieny B. 2012 Appl. Phys. Lett. **100** 142410.
[40] Akbulut S., Akbulut A., Özdemir M., Yildiz F. 2015 J. Magn. Magn. Mater. **390** 137.
[41] Sbiaa R., Shaw J. M., Nembach H. T., Al Bahri M., Ranjbar M., Åkerman J., Piramanayagam S. N. 2016 J. Phys. D: Appl. Phys. **49** 425002.
[42] Law W. C., Jin T. L., Zhu X. T., Nistala R. R., Thiyagarajah N., Seet C. S., Lew W. S. 2019 Magn J. Magn. Mater. **477** 124.
[43] Malinowski G., Boulle O., Klaui M. 2011 J. Phys. D: Appl. Phys. **44** 384005.